\documentclass[12pt]{iopart}

\usepackage{pdfpages}
\usepackage{graphicx}
\begin{document}

\title[]{Performance evaluation of coherent Ising machines against classical neural networks}

\author{Yoshitaka Haribara$^{1, 2}$, Hitoshi Ishikawa$^{3}$, Shoko Utsunomiya$^{4}$, Kazuyuki Aihara$^{1, 2}$ and Yoshihisa Yamamoto$^5$}

\address{$^1$ Department of Mathematical Informatics, University of Tokyo, 
Tokyo 113-8656, Japan}
\address{$^2$ Institute of Industrial Science, University of Tokyo, 
Tokyo 153-8505, Japan}
\address{$^3$ PEZY Computing, 
Tokyo 101-0052, Japan}
\address{$^4$ National Institute of Informatics, 
Tokyo 101-8430, Japan}
\address{$^5$ ImPACT Program, Japan Science and Technology Agency, 
Tokyo 102-0076, Japan}
\ead{haribara@sat.t.u-tokyo.ac.jp}
\vspace{10pt}
\begin{indented}
\item[]\today 
\end{indented}

\begin{abstract}
The coherent Ising machine is expected to find a near-optimal solution in various combinatorial optimization problems, which has been experimentally confirmed with optical parametric oscillators (OPOs) and a field programmable gate array (FPGA) circuit. The similar mathematical models were proposed three decades ago by J. J. Hopfield, et al. in the context of classical neural networks. In this article, we compare the computational performance of both models.
\end{abstract}

%
\vspace{2pc}
\noindent{\it Keywords}: Degenerate Optical Parametric Oscillator, Measurement Feedback System, Combinatorial Optimization Problems, Maximum Cut Problems

\submitto{Quantum Science and Technology (focus issue on ``Quantum coherent feedback and reservoir engineering")}

%

\section{Introduction}
In recent trends in semiconductor technologies, the Moore's law is slowing down mainly due to the limitation of micro-fabrication heat dissipation and communication bottleneck problems on a chip \cite{hennepatter2011computer,waldrop2016chips}. Many efforts to boost the processor performance have been made for parallelized architectures including GPU, other multi/many-core processors, and neuromorphic hardwares \cite{khan2008spinnaker}. An optics-based special purpose computer, which is named the coherent Ising machine (CIM), has been proposed to exploit a rapid physical convergence time for accelerating the solution search in hard optimization problems \cite{utsunomiya2011mapping}.

One of the well-known examples of combinatorial optimization problems is a maximum cut problem (MAX-CUT) on a graph, which is essentially equivalent to the Ising model in statistical mechanics \cite{garey2002computers}. It is a problem to find the largest cut in a given graph $G=(V,E)$, where the number of edges at the boundary of a partition of vertices into two subsets is maximized. The size of the cut is defined as the total weight of edges separated by the cut, i.e., edges which have each endpoints in the different sides of the cut. This objective function can be written as 
\begin{equation}\label{eq:CUT}
\mathrm{CUT}(x) = \sum_{1 \leq i < j \leq n} w_{ij} \frac{1 - x_i x_j}{2}, 
\end{equation}
where the graph order $n = |V|$ is the number of vertices, $w_{ij}$ is the weight of the edge $(i,j) \in E$ and $x_i = \pm 1$ is a binary value indicating which side of the cut the vertex $i \in V$ belongs to.

To implement the above problem on a physical system, the injection-locked lasers \cite{utsunomiya2011mapping} and the degenerate optical parametric oscillators (DOPOs) \cite{wang2013coherent} were proposed to use. With a series of experimental challenges, the implementation of the 
optical delay line based \cite{marandi2014network, takata201616} and measurement feedback based \cite{McMahon614, Inagaki603} coherent Ising machines has been realized. 

As a metaheuristic algorithm, the CIM can be interpreted as a mathematical model to solve combinatorial optimization problems using recurrently updated neurons with nonlinear activation function. From this point of view, there have been related and interesting approaches by mathematical models of the neurons (e.g., \cite{mcculloch1943logical, hodgkin1952quantitative}) and their networks (e.g., \cite{fukushima1980neocognitron}). Hopfield developed the optimization algorithm by using such neural networks \cite{
Hopfield01041982}. Then Hopfield and Tank extended it to the continuous valued model to improve the performance and applied it to the combinatorial optimization problems \cite{Hopfield01051984, hopfield1986computing}. Simulated annealing (SA) is proposed in the same period \cite{Kirkpatrick671}.

Here, we try to clarify the relative performance of our CIM against a family of classical neural network approaches and SA for combinatorial optimization problems especially for MAX-CUT. This paper is organized as follows. In \sref{sec:CIM}, we describe the basic concept and experimental configuration of CIM. In \sref{sec:model}, we describe models of neural network algorithm for the combinatorial optimization problems followed by \sref{sec:hard} to describe suitable hardware implementation. Then the numerical experiments are performed in \sref{sec:numerical}. We discuss the results and other possibilities of implementations in \sref{sec:discussion}. Finally, we conclude the paper in \sref{sec:conclusion}.

\section{Method}
\subsection{Coherent Ising machine (CIM)}\label{sec:CIM}
We intend to solve combinatorial optimization problems by mapping the cost function \eref{eq:CUT} to the energy of an Ising spin system. CIM is initially proposed as an injection-locked laser system \cite{utsunomiya2011mapping}, followed by the proposal using a degenerate optical parametric oscillator (DOPO) system \cite{wang2013coherent}. So far, several experimental machines are demonstrated with $n = 4, 16, 100, 2048$-pulse systems \cite{marandi2014network, takata201616, McMahon614, Inagaki603}. Since the original MAX-CUT has binary variables, we use a bistable optical device, DOPO at the output stage of computation, while an analog optical device, degenerate optical parametric amplifier (DOPA), at the solution search stage of computation. 

\begin{figure}
\begin{center}
\includegraphics[width=320pt]{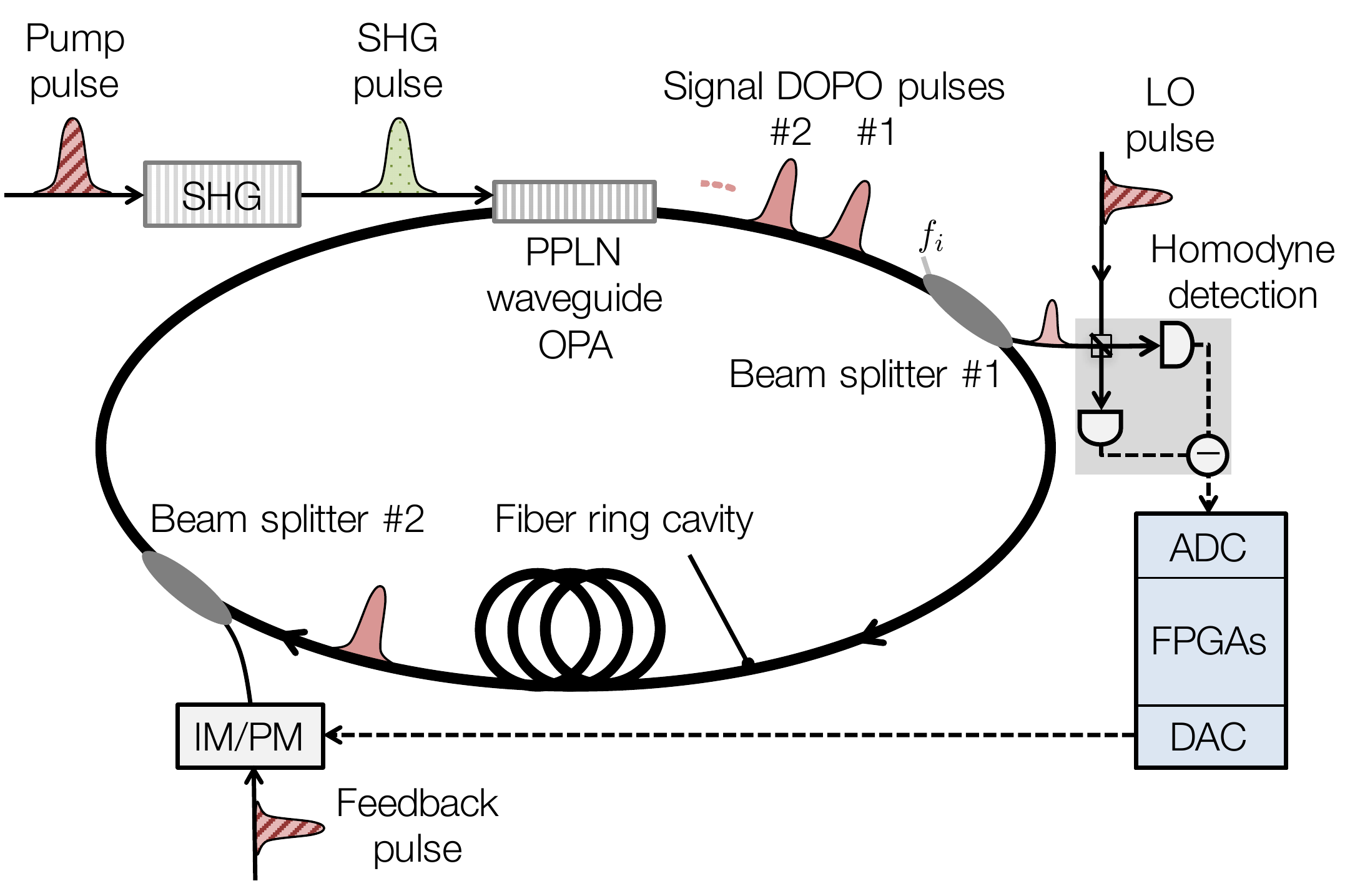}
\end{center}
\caption{\label{fig:CIM}Experimental schematic of a coherent Ising machine implemented on a fiber DOPO with an FPGA measurement feedback circuit.}
\end{figure}

\Fref{fig:CIM} depicts the schematic of the measurement feedback based CIM \cite{McMahon614, Inagaki603}. Here we describe the typical experimental configurations in \cite{Inagaki603}. The DOPO part consists of a 1 km optical fiber with an externally pumped periodically poled lithium niobate (PPLN) waveguide. The pulsed pump laser, at the 1 GHz repetition rate of 5000 times as the cavity circulation frequency, generates 5000 individual DOPO pulses in a single fiber ring cavity. A segment of them (2000 pulses) is used as the signal pulses for computation and the remaining portion (3000 pulses) is used for the cavity stabilization. 

The feedback circuit stores the interaction strength for each pair of DOPO pulses. A portion of optical pulse is picked-off by a beamsplitter (numbered as 1 in the \fref{fig:CIM}) and measured by balanced homodyne detectors. The measured values of DOPO pulse amplitudes are fed into an analog-digital converter (ADC), followed by FPGAs. Here, 1 GHz repetition rate of signal pulses is downclocked to 125 MHz (8 parallel) and the measured amplitudes $\tilde{c}_i$ are sliced into the digital signals of 5 bits. Then 2 FPGAs sum up the coupling effect from the other vertices (in the given topology) $\sum_j J_{ij} \tilde{c}_j$ for the $i$th pulse. The feedback pulse train is modulated in intensity and phase by this output electrical signal after a digital-analog converter (DAC). The feedback pulse is injected to the signal DOPO pulse running through the main fiber ring cavity via a beamsplitter \#2.

The DOPO is operated near the oscillation threshold by crossing the pump rate from below to above the threshold in the case of \cite{Inagaki603}. In the beginning, the DOPO is biased at below the threshold in which all phase configuration is established so as a superposition state and the quantum parallel search is implemented \cite{maruo2016truncated}. Then, the external pump rate (or the feedback) strength is gradually increased, and once the whole system reaches the oscillation threshold, it selects a particular phase configuration which corresponds to the near-optimal solution of the original optimization problem.

The dynamics of the CIM can be simulated by the quantum master equation. Instead of numerically integrating the master equation for the DOPO density operator, we can expand the density operator by the quasi-probability function in the phase space. One quasi-probability function need for this purpose is the positive $P(\alpha, \beta)$ representation in terms of the off-diagonal coherent state expansion, $\left| \alpha \right\rangle \left\langle \beta \right|$. The Fokker-Planck equation for $P(\alpha, \beta)$ is derived from the master equation and then the c-number stochastic differential equations for $\alpha$ and $\beta$ are obtained using the Ito calculus (see \cite{PhysRevA.92.043821} for detail). Another quasi-probability function used for this purpose is the truncated Wigner representation $W(\alpha)$ in terms of the Gaussian states. The corresponding c-number stochastic differential equations are derived in \cite{maruo2016truncated}. 

\subsection{Classical neural networks} \label{sec:model}
We describe in this section the classical neural network models to solve the same combinatorial optimization problems, three of which are summarized in the \tref{tab:nn}.

\begin{table}
\caption{\label{tab:nn}Classical neural-network approaches for combinatoral optimization problems.} 
\begin{indented}
\lineup
\item[]\begin{tabular}{l | l l}
\br                              
$$ & Deterministic & Stochastic \cr 
\mr
Binary & Derandomized Hopfield network (HN) & Simulated annealing (SA) \cr
Analog &Hopfield-Tank neural network (HTNN)& \cr 
\br
\end{tabular}
\end{indented}
\end{table}

\subsubsection{Derandomized Hopfield Network (HN)}
J. J. Hopfield implemented a classical neural network model solving combinatorial optimization problems in his 1982 paper \cite{Hopfield01041982}, which is referred to the Hopfield network (HN). The neuron in this model has the discrete output values $x_i = \pm 1$  with a simple majority voting update rule:
\begin{equation}
x_i \leftarrow \mathrm{sgn}(\sum_{j = 1}^nJ_{ij}x_j)
\end{equation}
which will execute asynchronously. The spin index $i$ is selected randomly in the original paper but we derandomized it to enhance the speed, i.e., the spin indices from $i=1$ to $i=n$ are updated sequentially. Simultaneous updates introduce the instability or periodic solution into the system. Since the update is local and deterministic, the system will converge into the nearest local minimum, which is determined by the initial state. Note that the model is originally proposed with $\{0,1\}$-binary neurons, but for comparison, we use equivalent $\{+1,-1\}$-valued neurons.

\subsubsection{Simulated Annealing (SA)}
While the HN will often get stacked at poor local minima, Kirkpatrick, et al. introduced a stochastic spin update strategy in simulated annealing (SA) algorithm to mimic thermal annealing \cite{Kirkpatrick671}. The probability of stochastic spin flip is governed by the Boltzmann factor in the Metropolis-Hastings procedure as follows:
\begin{equation}
P = \exp(- \Delta E_i / T), 
\end{equation}
even if the energy difference to flip the $i$th spin
\begin{equation}
\Delta E_i = 2 x_i \sum_{j=1}^n J_{ij}x_j
\end{equation}
makes the total energy increased, namely $\Delta E_i > 0$. The spin index $i$ is selected randomly while temperature $T$ is gradually decreased. 

\subsubsection{Hopfield-Tank Neural Network (HTNN)}
Hopfield and Tank proposed another neural network approach using an analog valued neuron $x_i \in [-1,1]$, which is referred to the Hopfield-Tank neural network (HTNN) \cite{hopfield1986computing}. The time evolution of the HTNN is described by ordinary differential equations (ODE):
\begin{equation}\label{eq:HTNN}
\frac{dx_i}{dt} = - \alpha x_i + \beta \sum_{j=1}^n J_{ij} f(x_j), 
\end{equation}
where $f(x)$ is a nonlinear sigmoid function. In this study, $\tanh(x)$ is used as $f(x)$. In the extremely high linear gain limit, i.e., when the slope of the sigmoid function around 0 is steep, this HTNN model becomes close to the HN model described above. The parameters in later section are optimized as the neuron decay rate $\alpha = 6$ and the synaptic connection strength $\beta = 0.1$ to achieve the best performance for the given MAX-CUT problems. The numerical integration of \eref{eq:HTNN} is performed by the Euler method with the discrete time step $\Delta t = 0.01$. 

\subsection{Hardware used for Implementation of classical neural networks}\label{sec:hard}
Here we describe the hardware configuration needed to implement the classical neural networks, which will be used in the benchmark section. Note that all codes are implemented with C++ \footnote{We used Ubuntu 16.04.4 with GCC 5.4.0 (CPU) and CentOS 7.1.1503 with GCC 4.8.3 (PEZY-SC)}.
\subsubsection{CPU (for SA and HN)}
SA and HN are iterative updating algorithms for discrete spins. We can achieve CPU implementation efficiently by SIMD bitwise operations in parallel\footnote{The code is available here. https://github.com/haribara/SA-complete-graph}. 
In this paper, we mainly used Intel Xeon E3-1225 v3 @ 3.2 GHz (Haswell architecture shipped in 2013). Note that the performance of SA is slightly improved from the previous paper, in which SA is run on an older processor (Intel Xeon X5650 @ 2.67 GHz Westmere architecture shipped in 2010) \cite{Inagaki603}. We did not use any accelerators for HN and SA in this study since it is already parallelized by SIMD operations in CPU and the cache hit rate is high enough as $98.8 \%$.

\subsubsection{MIMD many core processor PEZY-SC (for HTNN)}
\begin{figure}
\begin{center}
\includegraphics[width=420pt]{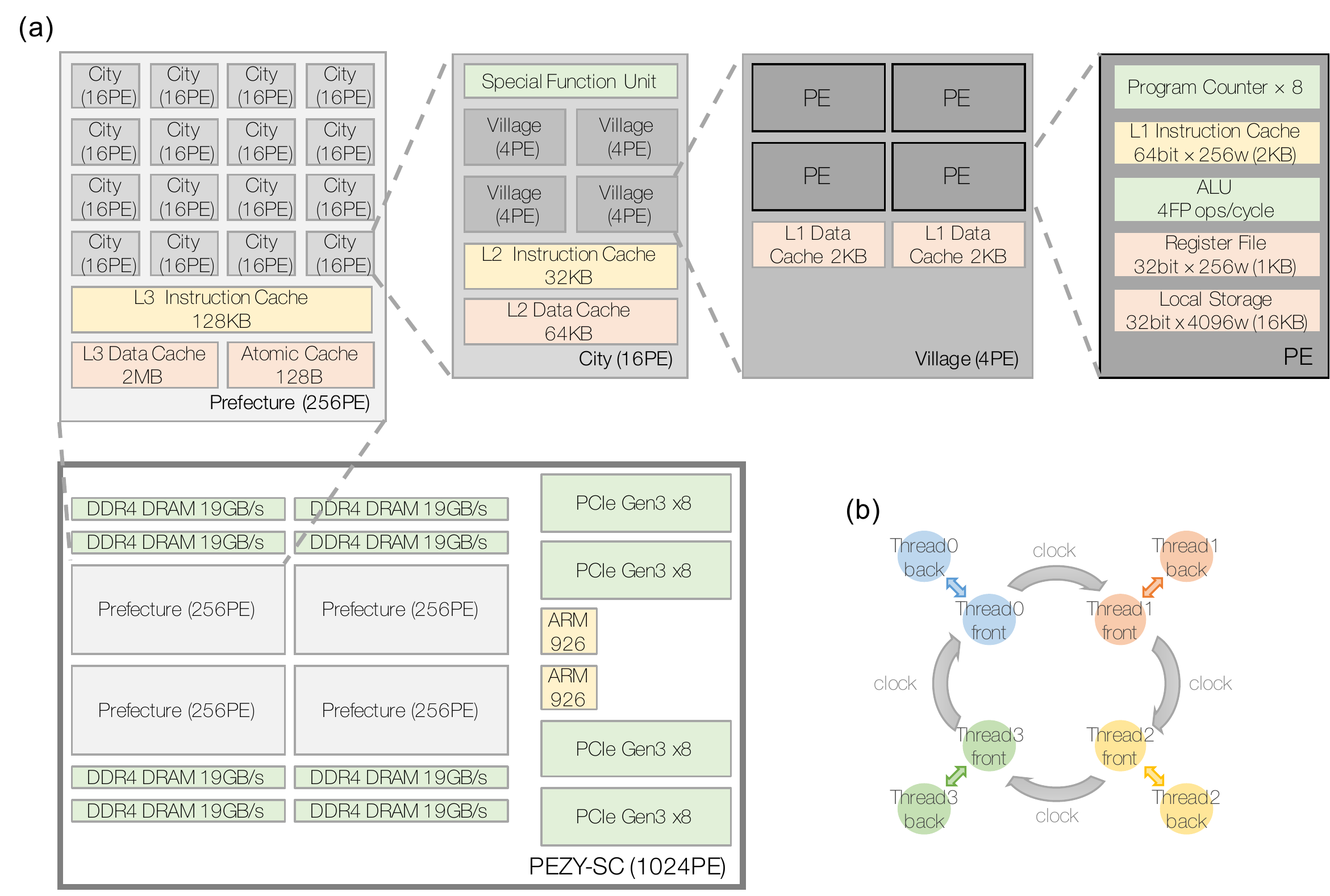}
\end{center}
\caption{\label{fig:PEZY}(A) The hierarchical architecture of a PEZY-SC many core processor. There are 1024 processing elements (PE) packed in a single PEZY-SC chip. (B) Each PE core handles 8 threads independently.}
\end{figure}

Since HTNN is based on ordinary differential equations (a continuous-valued continuous-time system) and requires floating-point arithmetic, it is better to parallelize by accelerators. We used a MIMD many core processor PEZY-SC @ 733 MHz with 1024 cores and 8192 threads on a chip (the architecture is shown in \fref{fig:PEZY}), which is set in Shoubu supercomputer at Riken (Japan). We parallelized matrix-vector multiplication and neuron updates in 8192-thread parallel. The coupling matrix is efficiently stored as a 1-bit matrix (since $J_{ij} = \pm 1$ has no empty entry) and neuronal states as floating points (32-bit float). Note that it was 1.4 times faster than storing matrix values in 32 bits. The benchmark of the hardware itself is shown in \ref{sec:appendix-PEZY}.

\section{Results}\label{sec:numerical}
\begin{figure}
\begin{center}
\includegraphics[width=320pt]{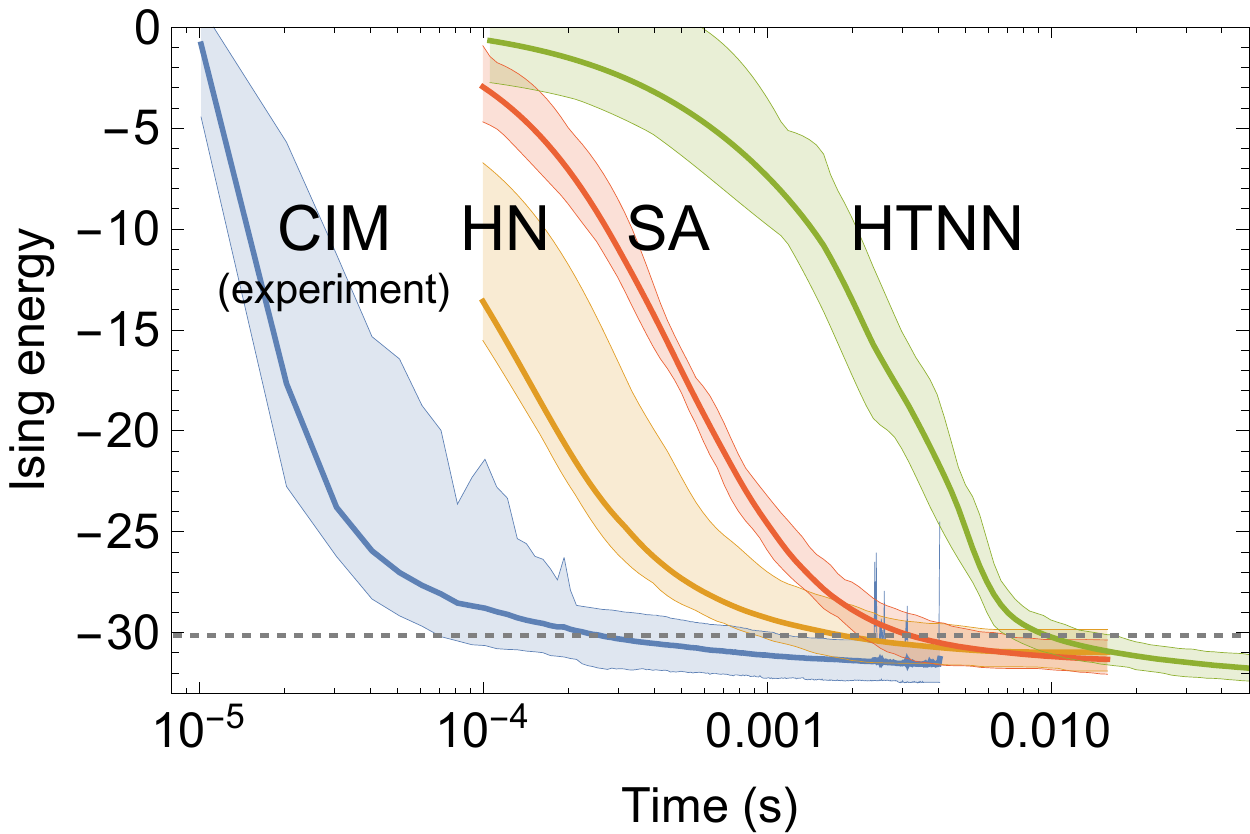}
\end{center}
\caption{\label{fig:N2000}Energy descent when solving a $\{+1,-1\}$-weighted $N=2000$ complete graph. Each thick line is the ensemble average of 100 trials (except for CIM experiment, which consists of 26 trials), while the lower and upper shaded error bars show the best and worst envelopes for each computational model. The gray dotted line is the target energy ($ = -60278/n$) which is obtained by the SDP relaxation algorithm \cite{goemans1995improved}.}
\end{figure}

We compared the performance of HN, SA, HTNN, and CIM by solving the MAX-CUT problems on a dense graph. The particular problem instance is a complete graph, in which all pair of $N = 2000$ vertices are connected and edges are weighted by $\{+1,-1\}$ in uniform distribution (the identical instance as in Ref. \cite{Inagaki603}). \Fref{fig:N2000} shows the performance on the complete graph, while the detailed computation time to target and the hardware configurations are summarized in \tref{tab:ttt}.

We ran 100 different trials for the same problem instance (except for CIM, which consists of 26 trials). Each solid line in the \fref{fig:N2000} indicates the ensemble average of all trials, while the lower and upper shaded lines indicate the best and worst case envelopes, respectively. Here, parameters for SA and HTNN are optimized to achieve the shortest computation time to the target which is obtained by the SDP relaxation algorithm \cite{goemans1995improved}.

The computation time to the SDP-produced target is shorter in the order of CIM, HN, SA, HTNN on the instance. The data from CIM are noisy due to experimental noise, but it can find better solutions than the target in all 26 trials. HN is faster than SA since HN can be regarded as a derandomized version of SA. Note that in the worst case, HN cannot reach the target (it fails 3 times in 100 trials as it can be seen partly in the worst case in \fref{fig:N2000}). It can be understand that HTNN performs much slower than HN/SA since it solves ODEs which deals with the analog variables. Note that HTNN achieves lower energy than SA in \fref{fig:N2000} but the performance of SA heavily depends on temperature scheduling. We optimized to reach the target shorter but slower scheduling ends up lower energy generally.

\begin{table}
\caption{\label{tab:ttt}Time to target and hardware configurations. The best (shortest) time to reach the target value and the time to cross the ensemble averaged line are listed in the upper table. Note that the target value is obtained by the SDP relaxation algorithm which has performance guaranty to the $87\%$ of the optimal value.} 
\begin{indented}
\lineup
\item[]\begin{tabular}{l | l  l | l}
\br                              
$$ & Best (ms) & Average (ms) & Hardware \cr 
\mr
CIM & 0.071 & 0.264 & fiber DOPO+2FPGAs\cr
HN & 0.924 & 1.84 & CPU\cr 
SA & 2.10 & 3.20 & CPU\cr
HTNN & 7.04 & 9.67 & PEZY-SC\cr
\br
\end{tabular}
\end{indented}

\begin{indented}
\lineup
\item[]\begin{tabular}{l | l  l | l l l}
\br                              
$$Hardware & Model & Clock and Architecture & Release date \cr 
\mr
FPGA & Xilinx Virtex-7 VX690T& 125 MHz 
693k logic cells & 2010 \cr
CPU & Intel Xeon E3-1225 v3 & 3.2 GHz, Haswell & 2013\cr
MIMD many core & PEZY-SC & 733 MHz, 1024 core & 2014\cr
\br
\end{tabular}
\end{indented}

\end{table}

\section{Discussion} \label{sec:discussion}
In this section, we will add the two discussions to justify the above conclusions, 
\begin{itemize}
\item Validity for the hardware selection.
\item Optimization for PEZY-SC implementation for HTNN.
\end{itemize}

\subsection{Validity for the hardware selection}
HTNN is apparently efficient on PEZY-SC than on CPU, which is shown in \ref{sec:appendix-PEZY}. On the other hand, we did not use any accelerator for HN and SA in this study. This is because we do not expect significant speed-up by naive implementation since they are already parallelized by SIMD operations in CPU and the cache hit rate is so high as $98.8 \%$ (measured by perf command in Linux). Generally, asynchronous update in HN/SA seems to be not suitable for parallel implementation. 

\subsection{Optimization for PEZY-SC implementation for HTNN}
We tried to optimize the implementation by storing matrix data efficiently. Since the given adjacency matrix has only the 1-bit entry ($J_{ij} = \pm 1$), we packed each value in 1 bit. This contributes the 1.4 times speed-up than having a 32-bit float matrix. But putting the data in local memory does not contribute to significant speed-up since its bottleneck in computation is not in memory transfer. There is a possibility of speed-up if we replace the multiplication by the selector. 
Rather, it is possible to scale out for parallel distributed processing by using multiple PEZY-SC chips in Soubu supercomputer, especially when the problem size is larger. 

\section{Conclusion} \label{sec:conclusion}
In this paper we compared the performance of the CIM implemented on DOPOs and FPGAs against the family of classical neural-network-based algorithms: HN, SA and HTNN. To accelerate the performance of the classical neural networks, HN and SA are implemented on CPU with bit operations and HTNN is implemented on a many core processor PEZY-SC. It is shown that the CIM can achieve faster computational time than HN (13.0 times for the best case and 6.97 times for the average), SA (29.6 times for the best case and 12.1 times for the average) and HTNN (99.2 times for the best case and 36.7 times for the average). 
\ack
The authors thank 
H. Takesue and T. Inagaki for providing the experimental data, 
K. Kawarabayashi, S. Tamate and T. Sonobe for accelerating SA implementation, 
Y. Saito for comments on the FPGA configuration, 
T. Leleu and M. Oku for general discussion.
We used a super computer Shoubu in Riken (Saitama, Japan) to benchmark on PEZY-SC. 
This research is supported by the Impulsing Paradigm Change Through Disruptive Technologies (ImPACT) Program of the Council of Science, Technology and Innovation (Cabinet Office, Government of Japan). 

\appendix
\section{}\label{sec:appendix}
\subsection{Processor performance of PEZY-SC}\label{sec:appendix-PEZY}
We show the elapsed time for the CIM simulation by the following c-number stochastic differential equations
\cite{maruo2016truncated, haribara2016computational}
\begin{eqnarray}
&&dc_{i}=[(p- c_i^2- s_i^2)c_i ]\, dt +\frac{1}{A_\mathrm{s}}\sqrt{c_i^2+s_i^2+\frac{1}{2}}dW_{\mathrm{c}i} \label{eq:gain},\\
&&c_{i}(t+\Delta t) \mapsto \sqrt{1-T_\mathrm{mes}}c_{i}(t) + \sqrt{T_\mathrm{mes}}\frac{f_i}{A_\mathrm{s}} \label{eq:loss},\\ 
&&c_{i}(t+\Delta t) \mapsto \sqrt{1-T_\mathrm{inj}}c_{i}(t) + \sqrt{T_\mathrm{inj}}\xi\sum_{j=1}^n J_{ij}\tilde{c}_j \label{eq:coupling}, 
\end{eqnarray}
implemented on different processor configurations. Note that the simulation by the Langevin equation is simplified version and we employed this model to contrast the processor performance (detailed simulation will be reported in \cite{shoji2017positive, yamamura2017density}). Here, CPU indicates serialized calculation on a single thread, CPU+GPU indicates matrix-vector multiplication is off-roaded to GPU while other part of differential equation is calculated on the same processor as CPU, PEZY-SC indicates that all processes are paralellized. We conclude that is it preferable to implement HTNN on PEZY-SC than CPU or GPU which we listed in the \tref{tab:acc} since the Langevin equations are similar to ODEs of HTNN except for random number generation. 

\begin{table}
\caption{\label{tab:acc}Accelerator configurations for parallel implementation of the c-number stochastic differential equations in \fref{fig:sim}.} 
\begin{indented}
\lineup
\item[]\begin{tabular}{l | l  l | l l l}
\br                              
$$Hardware & Model & Clock and Architecture & Release date \cr 
\mr
CPU & Intel Xeon W3530 & 2.80 GHz, Nehalem-WS & 2010\cr
GPU & NVIDIA Tesla C2075 & 1.15 GHz, Fermi & 2011\cr
MIMD many core & PEZY-SC & 733 MHz, 1024 core & 2014\cr
\br
\end{tabular}
\end{indented}
\end{table}

\begin{figure}
\begin{center}
\includegraphics[width=320pt]{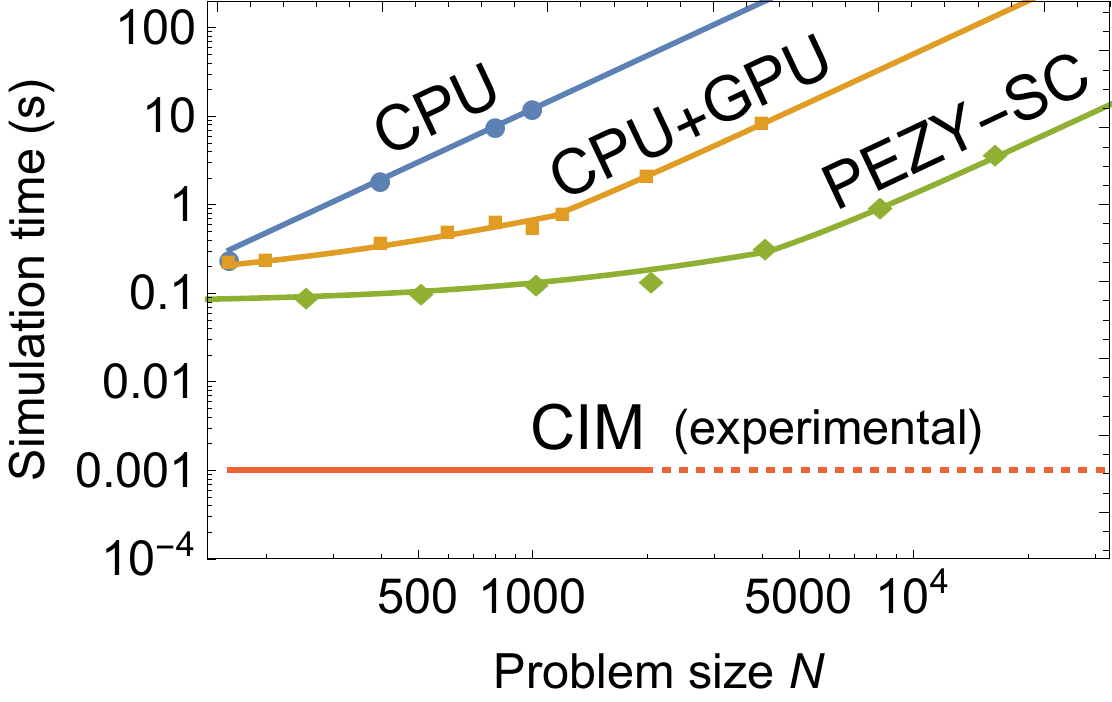}
\end{center}
\caption{\label{fig:sim}Simulation time for the c-number stochastic differential equations (200 round trips on complete graphs).}
\end{figure}
\section*{References}
\bibliography{IOP-QST}
\end{document}